\documentclass{article}

\usepackage{arxiv}

\usepackage[utf8]{inputenc} 
\usepackage[T1]{fontenc}    
\usepackage{hyperref}       
\usepackage{url}            
\usepackage{booktabs}       
\usepackage{amsfonts}       
\usepackage{nicefrac}       
\usepackage{microtype}      
\usepackage{lipsum}
\usepackage{tikz}
\usepackage{quantikz}
\usepackage{hyperref}
\usepackage{blochsphere}
\usepackage{subcaption}
\usetikzlibrary{calc}
\usetikzlibrary{arrows.meta}
\usetikzlibrary{angles}
\usetikzlibrary{quotes}
\usepackage{graphicx}

\graphicspath{ {./images/} }

\title{Performing a Quantum Subroutine Using Cluster States}

\author{
Sakibul Islam Rayhan \\
Electrical and Electronic Engineering\\
Shahjalal University of Science and Technology\\
Sylhet, Bangladesh \\
\texttt{sakibul56@student.sust.edu} \\
\And  
Karima Hassan \\
Electrical and Electronic Engineering\\
Shahjalal University of Science and Technology\\
Sylhet, Bangladesh \\
\texttt{karima75@student.sust.edu} \\
\And
Md Rasedujjaman \\
Electrical and Electronic Engineering\\
Shahjalal University of Science and Technology\\
Sylhet, Bangladesh \\
\texttt{mrased-eee@sust.edu} \\
\And
Md Mohsinur Rahman Adnan \\
Electrical and Electronic Engineering \\
Bangladesh University of Engineering and Technology\\
Dhaka, Bangladesh\\
\texttt{mmradnan@eee.buet.ac.bd} 
}

\begin{document}
\maketitle
\begin{abstract}
Quantum subroutines, like their classical counterparts, serve as essential components of larger algorithms. They can be realized using gate based architectures or through measurement based quantum computation (MBQC), where computation proceeds via adaptive measurements on highly entangled cluster states. In this work, we implement a phase oracle as a reusable function within the MBQC framework using experimentally feasible cluster states. Unlike prior demonstrations of Grover’s algorithm in MBQC which either omit explicit oracle construction. Our approach provides a transparent formulation of the oracle subroutine, including its measurement pattern and byproduct correlation, density matrix simulations also the corresponding MBQC measurement circuits.
\end{abstract}

\keywords{Measurement Based Quantum Computing \and Cluster States \and Quantum Subroutine \and Phase Oracle}

\section{Introduction}
Subroutines are the basic building blocks of computer programs. As the classical algorithm got a subroutine, so as the quantum algorithm. Quantum subroutine is a modular block of a quantum algorithm, reusable like classical functions \cite{qsub1}. Past quantum subroutines are mostly for gate based quantum computing \cite{qsub1, qsub2, qsub3}. Building a subroutine with cluster state, which basically relies on adaptive measurements on pre-prepared entangled cluster states. This type of computation is called measurement based quantum computation (MBQC) or "one way" quantum computation \cite{MBQC1}.  MBQC or one-way quantum computing uses a method that first prepares an entangled resource state, usually a cluster state or graph state \cite{cstate}, then performs single qubit measurements on it. It is ”one-way” because the resource state is destroyed by the measurement \cite{MBQC2}. We choose MBQC because it avoids noisy in line entangling gates by generating entanglement off line, making implementation simpler and keeping optical circuit depth constant, which minimizes losses in photonic systems \cite{opticalmbqc1}\cite{opticalmbqc2}. 

Prior studies for MBQC include experimentation with Grover’s Algorithm \cite{MBQC1}, Variational Quantum Eigensolver \cite{mbqcalgo}, Quantum Approximate Optimization Algorithm \cite{mbqcalgo2}. For our work, we choose grover’s phase oracle, which will work as a simple subroutine for a larger algorithm. Here we present a comprehensive study that establishes the mathematical construction of a two qubit Grover's phase oracle within the framework of MBQC. We provide an explicit cluster state representation of the oracle, accompanied by density matrix simulations and MBQC circuit with measurement sequence with byproduct correlation.

\section{Background}

\subsection{Cluster States}

A highly entangled multi qubit resource state is called cluster state or a more general graph state \cite{cstate}. In quantum information theory, cluster states form an important class of multipartite entangled states defined on
qubits arranged in a lattice structure \cite{ccluster2}. A cluster $C$ is a connected subset of a $d$-dimensional lattice, and a cluster state is a pure quantum state associated with the qubits located on the sites of $C$. These states may be generated through Ising-type interactions applied across the lattice.

Formally, a cluster state $|\phi_{\{k\}}\rangle_{C}$ is defined as the simultaneous eigenstate of a family of correlation operators $K^{(a)}$, satisfying
\begin{equation}
    K^{(a)} |\phi_{\{k\}}\rangle_C = (-1)^{k_a} |\phi_{\{k\}}\rangle_C,
\label{eq:cluster_eigen}
\end{equation}
where $\{ k_a \in \{0,1\} \,|\, a \in C \}$ specifies the instance of the cluster state. Each correlation operator takes the form
\begin{equation}
    K^{(a)} = \sigma_x^{(a)} \bigotimes_{b \in N(a)} \sigma_z^{(b)},
\label{eq:cluster_operator}
\end{equation}
with $\sigma_x$ and $\sigma_z$ the Pauli operators, and $N(a)$ denoting the neighborhood of site $a$ \cite{MBQC2}. 





\subsection{Measurement Based Quantum Computing}

Measurement based quantum computation (MBQC), also known as the one way quantum computer, is a computational
model in which a highly entangled multi qubit resource state is prepared prior to computation. The computation then proceeds solely by performing adaptive single qubit measurements on this state. The term “one way” reflects the irreversible consumption of the resource state as the measurements progress \cite{MBQC1}. Information propagation through a cluster state closely resembles quantum teleportation \cite{qteleport}. MBQC fundamentally exploits this mechanism, allowing quantum information to be transferred and transformed through successive single qubit measurements on the entangled resource. A crucial component of this process is classical feedforward, in which the measurement bases for later qubits are adjusted according to earlier measurement outcomes to ensure correct logical evolution despite inherent measurement randomness \cite{cfeed1}. The measurement pattern can be represented using the measurement sequence and byproduct correlations \cite{mpattern}. Also, MBQC circuit can be realized just like gate based quantum computation \cite{realization}. \\

\subsection{Phase Oracle \& Subroutine}

Grover’s algorithm, also known as the quantum search algorithm, includes an oracle that is a unitary operation that marks the correct solution to a search problem by applying a negative phase to their corresponding quantum states, while leaving all other states unchanged in phase  \cite{grov}.  Grover’s oracle is a phase oracle that marks a desired computational basis state by introducing a relative phase flip. Formally, let the Boolean function \( f(x) \) satisfy \( f(x) = 1 \) for the marked element \( x = w \) and \( f(x) = 0 \) for all \( x \neq w \). The action of the oracle \( U_w \) on the computational basis states is then defined as

\begin{equation}
U_w \ket{x} = (-1)^{f(x)} \ket{x},
\end{equation}

which implies that only the marked state acquires a negative phase.
The oracle operator can be written explicitly as $U_w = I - 2 \ket{w}\bra{w}.$ where \( I \) denotes the identity operator. Acting on the marked state \( \ket{w} \), the oracle yields

\begin{equation}
 \begin{split}
U_w \ket{w} 
= (I - 2\ket{w}\bra{w}) \ket{w} 
= \ket{w} - 2\ket{w} \langle w|w \rangle 
= \ket{w} - 2\ket{w} 
= -\ket{w} 
 \end{split}
\end{equation}

For all unmarked states \( \ket{x} \) with \( x \neq w \), the oracle acts trivially as \( U_w \ket{x} = \ket{x} \). Thus, the operator \( U_w \) selectively flips the phase of the target state while leaving all other basis states unchanged. This phase marking is the key mechanism that enables Grover’s algorithm to amplify the probability amplitude of the desired state through subsequent diffusion operations. There are different types of oracle we can find in different algorithms\cite{oracle1, oracle2, oracle3, oracle4, oracle5, oracle6}. 

These oracles encode classical or structured functions into unitary transformations and serve as black box , which allow an algorithm designer to build up an algorithm in modular block which can be called as subroutine that a quantum algorithms query to extract global properties.

\begin{figure}
 \centering
    \includegraphics[width=6in]{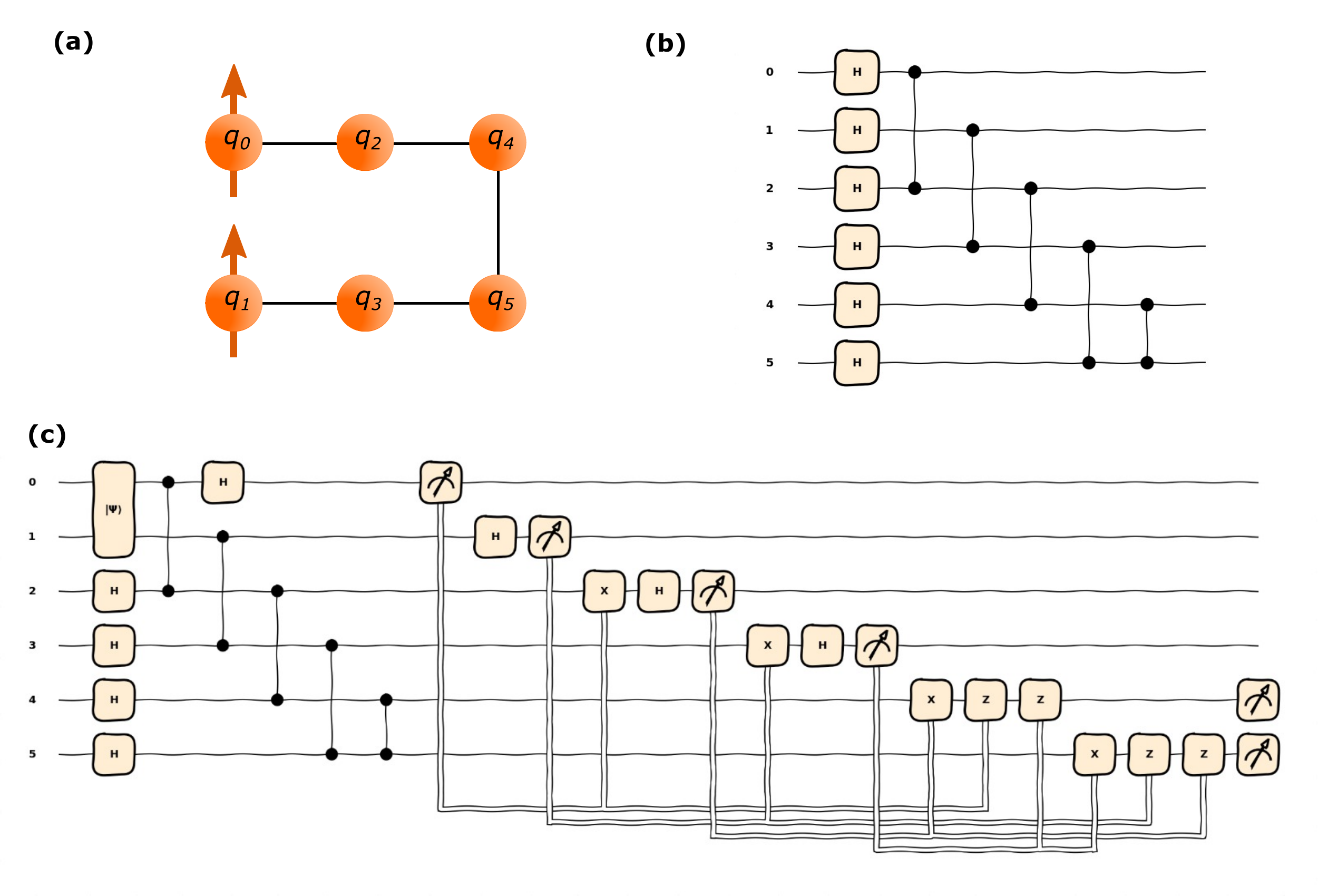} 
    \caption{Illustration of cluster states for the controlled‑Z (CZ) gate like functionality: (a) schematic representation of the CZ cluster state with six qubit arranged in $2\times3$ grid where $q_{0}, q_{1}$ is the input qubit and $q_{4}, q_{5}$ is the output qubit (b) Quantum circuit construction for generating an $6$-qubit cluster state. Each qubit is first initialized in the $\ket{+}$ state by applying a Hadamard gate to $\ket{0}$. Controlled-$Z$ ($CZ$) gates are then applied sequentially between neighboring qubits to entangle them. (c) Realization of the complete $CZ$-gate cluster state MBQC circuit. The input qubits $q_{0}$ and $q_{1}$ are initialized in the $XY(0)$ plane, while the remaining qubits follow the prescribed measurement pattern to implement the computation.}
    \label{fig:cluster_state} 
\end{figure}


\section{Methodology}
\subsection{Cluster State Preparation}

To implement the CZ-gate in MBQC, we require six qubits in an entangled state: two input qubits $|\psi \rangle$ and $|\phi 
\rangle$, along with four additional qubits prepared in the $|+ \rangle$ state. The six qubits are arranged in a 2 × 3 grid, with the first row containing qubits q0 (input: $|\psi\rangle$), q2, and q4, and the second row containing qubits q1 (input: $|\phi\rangle$), q3, and q5 \cite{LeeChung2025}. For this six qubit cluster state used in our MBQC construction, the expanded stabilizer generators derived from Equations \ref{eq:cluster_eigen}, \ref{eq:cluster_operator} are listed in Table \ref{tab:cluster_stabilizers}. These stabilizers define the entanglement structure of the cluster and ensure that the state remains invariant under the corresponding correlation operators. For that the state vector equation of the cluster state is shown in Equation \ref{maineq}. To physically realize this entangled cluster state, we apply Hadamard gates to each qubit followed by CZ gates between the pairs specified in the Table \ref{tab:cluster_stabilizers}.

\begin{table}[h!]
\centering
\begin{tabular}{c c}
\toprule
Qubit & Stabilizer $k^{(q_i)}$ \\
\midrule
$q_0$ & $\sigma_x^{(q_0)} \sigma_z^{(q_2)}$ \\
$q_1$ & $\sigma_x^{(q_1)} \sigma_z^{(q_3)}$ \\
$q_2$ & $\sigma_x^{(q_2)} \sigma_z^{(q_0)} \sigma_z^{(q_4)}$ \\
$q_3$ & $\sigma_x^{(q_3)} \sigma_z^{(q_1)} \sigma_z^{(q_5)}$ \\
$q_4$ & $\sigma_x^{(q_4)} \sigma_z^{(q_2)} \sigma_z^{(q_5)}$ \\
$q_5$ & $\sigma_x^{(q_5)} \sigma_z^{(q_3)} \sigma_z^{(q_4)}$ \\
\bottomrule
\end{tabular}

\vspace{0.3cm} 

\caption{Stabilizer generators defining the six-qubit cluster state. Each stabilizer corresponds to the product of a Pauli-$X$ operator on the given qubit and Pauli-$Z$ operators on its neighbors, reflecting the entanglement structure of the cluster.}
\label{tab:cluster_stabilizers}
\end{table}

\begin{equation}
    \begin{split}
        | \psi \rangle = \frac{1}{8} \ket{000000} + \frac{1}{8} \ket{000001} 
        + \frac{1}{8} \ket{000010} + \frac{1}{8} \ket{000011} + \frac{1}{8} \ket{000100}
        -  \frac{1}{8} \ket{000101} + ... - \frac{1}{8} \ket{111011} \\
        - \frac{1}{8} \ket{111100} + \frac{1}{8} \ket{111101} + \frac{1}{8} \ket{111110} - \frac{1}{8} \ket{111111}
    \end{split}
    \label{maineq}
\end{equation}

An illustration is given in Fig \ref{fig:cluster_state} a and b demonstrating the cluster state and circuit to built the cluster state.

\subsection{Measurement Pattern \& Byproduct Correlation}


To realize a phase oracle that marks the computational basis states \(\ket{00},\ket{01},\ket{10},\ket{11}\) within the MBQC paradigm, one must choose an appropriate measurement pattern on a suitable cluster (graph) state. the CZ-gate can be applied to the two input qubits by measuring 
$q_{0}, q_{1}, q_{2}$ and $q_{3}$ sequentially along the X basis. Considering the CZ gate interation the measurement sequence can be written like in the Equation \ref{sequence} \cite{LeeChung2025}.

\begin{equation}
    \begin{split}
        (X_{4}^{s_{2}}Z_{4}^{s_{3}}Z_{4}^{s_{0}} \otimes X_{5}^{s_{3}}Z_{5}^{s_{2}}Z_{5}^{s_{1}}) CZ (|\psi \rangle_{4} \otimes \ket{\phi}_{5})
    \end{split}
    \label{sequence}
\end{equation}

\begin{figure}
\centering
    \includegraphics[width=5in]{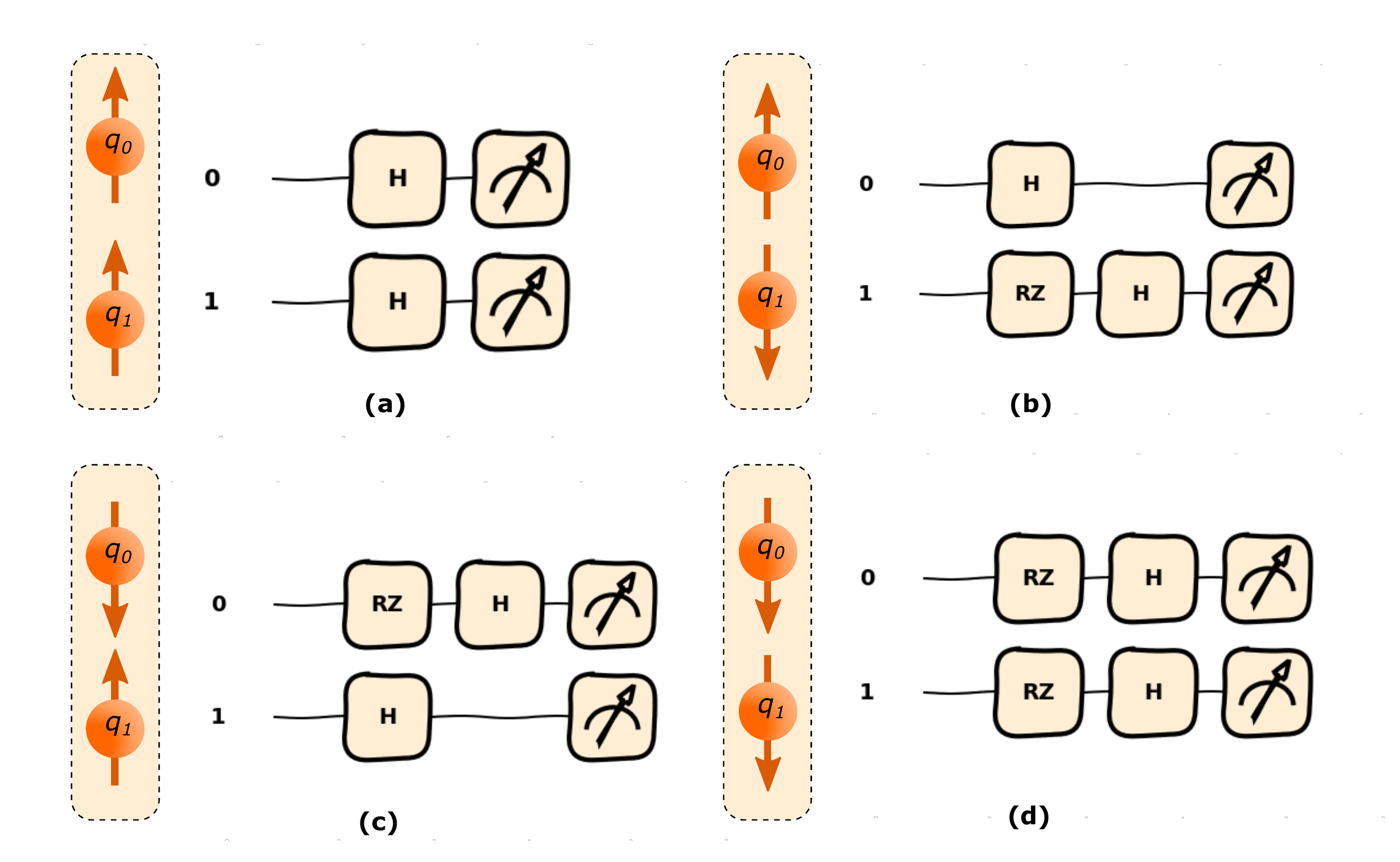} 
    \caption{Oracle construction based on input qubits $q_{0}$ and $q_{1}$, with rotations defined in the $XY$ plane: 
(a) choosing $XY(0)$ plane for both $q_{0}$ and $q_{1}$ to implement a phase shift on state $\ket{11}$; 
(b) choosing $XY(0)$ and $XY(\pi)$ for $q_{0}$ and $q_{1}$ to implement a phase shift on $\ket{10}$; 
(c) choosing $XY(\pi)$ and $XY(0)$ for $q_{0}$ and $q_{1}$ to implement a phase shift on $\ket{01}$; 
(d) choosing $XY(\pi)$ plane for both $q_{0}$ and $q_{1}$ to implement a phase shift on $\ket{00}$.}
    \label{fig:rotations} 
\end{figure}

By encoding information in qubit $q_{0}$ and $q_{1}$ with appropriate rotation one can make an oracle that will change the phase of the desired quantum state. In addition to it, an illustration of MBQC circuit has been added in Fig \ref{fig:cluster_state} c with proper measurement pattern and byproduct correlation. Choosing a suitable measurement basis for $q_{0}$ and $q_{1}$ is important for the phase oracle which is shown in Fig \ref{fig:rotations}.



\subsection{Density Matrix Extraction}

For the claim of oracle density matrix provides a convenient way to track how a quantum operation transforms the state, and this perspective is equally important when comparing gate based quantum computation with measurement based quantum computation (MBQC). In the MBQC model, single qubit measurements together with their associated byproduct corrections must reproduce the same state transformation as the corresponding unitary gate in the circuit model. Consider an arbitrary single qubit state $|\psi\rangle = \alpha |0\rangle + \beta |1\rangle$ whose density matrix is given by $\rho_{\text{in}} = |\psi\rangle\langle\psi| 
= 
\begin{pmatrix}
|\alpha|^{2} & \alpha\beta^{*} \\
\alpha^{*}\beta & |\beta|^{2}
\end{pmatrix}.$ Applying the phase rotation gate $R_{Z}(\theta) \ket{\psi} = 
\begin{pmatrix}
e^{-i\theta/2} & 0 \\
0 & e^{\,i\theta/2}
\end{pmatrix}$ transforms the state to $
    |\psi'\rangle = R_{Z}(\theta)|\psi\rangle
= \alpha e^{-i\theta/2}|0\rangle + \beta e^{\,i\theta/2}|1\rangle $. The corresponding output density matrix becomes the following Equation. 

\begin{equation}
    \rho'_{\text{out}} = |\psi'\rangle\langle\psi'|
=
\begin{pmatrix}
|\alpha|^{2} & \alpha\beta^{*} e^{-i\theta} \\
\alpha^{*}\beta e^{\,i\theta} & |\beta|^{2}
\end{pmatrix}
\label{density_m}
\end{equation}

This provides the means by which we can test the functionality of our MBQC oracle.

\section{Result \& Discussion}
  
\begin{center}
\begin{figure}[h!]
    \centering
    \includegraphics[width=0.6\textwidth]{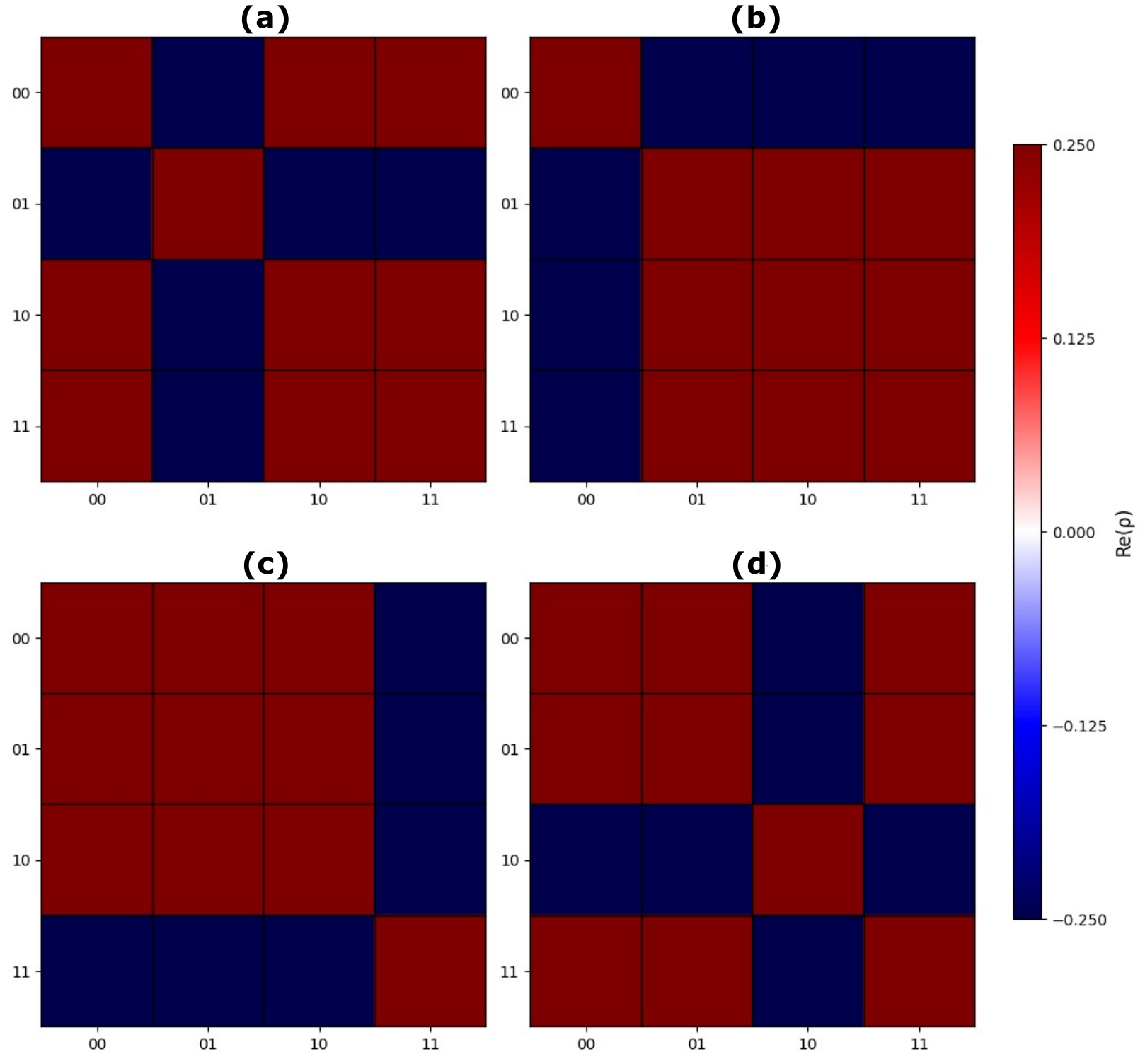} 
    \caption{Density matrix verification for the initial state $\ket{+}\otimes\ket{-}$, with outcomes shown for different measurement-based quantum computation (MBQC) oracles. The real part of the amplitude is displayed, with maximum values of $0.25$ and $-0.25$: (a) density matrix corresponding to the $\ket{11}$ marked oracle; (b) density matrix corresponding to the $\ket{10}$ marked oracle; (c) density matrix corresponding to the $\ket{01}$ marked oracle; (d) density matrix corresponding to the $\ket{00}$ marked oracle.}
    \label{fig:oracle_density_matrix}
\end{figure}
\end{center}

To assess the correctness of our measurement based quantum computation (MBQC) implementation, which is shown in Figure \ref{fig:rotations}. The phase oracle, we evaluate the output density matrices for each marked basis state and compare them with those obtained from the corresponding gate-based circuit model. A direct comparison using computational basis states such as $\ket{00}$, $\ket{01}$, $\ket{10}$, and $\ket{11}$ is problematic. After the application of a controlled $Z$ gate, several of these states yield identical density matrices, making it impossible to distinguish different oracle actions. To overcome this issue, we choose the initial state $\ket{+}\otimes\ket{-}$, which is sensitive to phase changes induced by the oracle. This allows us to distinguish the four possible marked elements. Figure~\ref{fig:oracle_density_matrix} shows the real part of the density matrices produced by the MBQC circuit for all four oracle configurations. Each subplot captures the amplitude distribution and correlation structure of the final two qubit state after the prescribed measurement pattern and associated byproduct corrections have been applied.

 Across all four cases, the density matrices generated by MBQC exhibit the same structural features as their gate based counterparts. In particular, the real part of each density matrix reflects the expected phase inversion on the marked state, while all unmarked components remain unchanged. This agreement demonstrates that the measurement pattern, together with the analytically derived byproduct operators, successfully implements the intended oracle transformation on the underlying cluster state. The consistency observed across all oracle configurations further validates our comparative methodology. By directly analyzing the output density matrices, we establish a clear operational equivalence between the gate based and MBQC realizations of the oracle. These results highlight the robustness of MBQC in reproducing circuit level behavior when supported by an appropriate entanglement structure and adaptively chosen measurement bases.

\section{Conclusion}

The present work establishes a unified and rigorous framework for comparing gate based subroutine with their cluster state based subroutine.
Although the analysis presented in this work focuses on small qubit configurations for reasons of physical feasibility, the framework itself is naturally extensible. With suitably designed measurement patterns and cluster geometries, the approach can be generalized to larger systems and more sophisticated computational tasks. This creates opportunities
for future investigations involving multi qubit MBQC constructions, fault tolerant extensions, and comparisons that incorporate higher dimensional cluster states. In this way, the framework offers a foundation for deeper insights into the operational equivalence of different models of quantum computation. It provides a structured method for analyzing,
validating, and extending MBQC implementations, contributing both to theoretical understanding and to the practical
development of scalable quantum technologies.

\section*{Data Availability}

The data and code supporting the findings of this study are publicly available in the \href{https://github.com/Rayhan740/Performing-a-Quantum-Subroutine-Using-Cluster-States.git}{GitHub repository}.

\section*{Acknowledgments}

We sincerely thank to Md Sakibul Islam, Department of Electrical and Computer Engineering, University of Central Florida for his guidance and insightful feedback throughout this work also thanks to KM Shariat Ullah and Ananta Kumar Das for their help with the illustrations.


\bibliographystyle{plain}

\section*{Appendix}

\renewcommand{\thesubsection}{(\alph{subsection})}

\subsection{Explicit Construction of One Dimensional Cluster States}

Cluster states are highly entangled multi-qubit states defined on a graph (or lattice), where qubits occupy the vertices and entanglement is generated along the edges via Ising-type interactions. For a connected subset $C$ of a $d$-dimensional lattice, a cluster state $|\phi_{\{k\}}\rangle_C$ is uniquely specified as the simultaneous eigenstate of a complete set of commuting stabilizer (correlation) operators
\begin{equation}
K^{(a)} = \sigma_x^{(a)} \bigotimes_{b \in N(a)} \sigma_z^{(b)},
\end{equation}
satisfying
\begin{equation}
K^{(a)} |\phi_{\{k\}}\rangle_C = (-1)^{k_a} |\phi_{\{k\}}\rangle_C,
\end{equation}
where $N(a)$ denotes the neighborhood of site $a$ and $k_a \in \{0,1\}$.  
In this appendix, we explicitly construct one-dimensional cluster states for $n=2,3,$ and $4$ qubits in the case $k_a=0$ for all $a$.

\subsection*{Two Qubit Cluster State ($n=2$)}

For a two qubit linear cluster, the stabilizer operators are,
\begin{equation*}
\begin{aligned}
K^{(1)} &= \sigma_x^{(1)} \otimes \sigma_z^{(2)}, \\
K^{(2)} &= \sigma_x^{(2)} \otimes \sigma_z^{(1)} .
\end{aligned}
\end{equation*}

We consider a general two-qubit state,
\begin{equation*}
|\phi\rangle_C = \alpha |00\rangle + \beta |01\rangle + \gamma |10\rangle + \delta |11\rangle .
\end{equation*}

Imposing the stabilizer conditions $K^{(1)}|\phi\rangle_C = |\phi\rangle_C$ and $K^{(2)}|\phi\rangle_C = |\phi\rangle_C$ yields the constraints,
\begin{equation*}
\alpha = \beta = \gamma = -\delta .
\end{equation*}

Normalization gives $4|\alpha|^2 = 1$, hence $\alpha = \frac{1}{2}$. The resulting cluster state is,
\begin{equation*}
|\phi\rangle_C = \frac{1}{2} (|00\rangle + |01\rangle + |10\rangle - |11\rangle),
\end{equation*}
which can be equivalently written as,
\begin{equation}
|\phi\rangle_C = \frac{1}{\sqrt{2}} (|0+\rangle + |1-\rangle).
\end{equation}

\subsection*{Three Qubit Cluster State ($n=3$)}

For a three qubit linear cluster, the stabilizer operators are,
\begin{equation*}
\begin{aligned}
K^{(1)} &= \sigma_x^{(1)} \otimes \sigma_z^{(2)} \otimes I^{(3)}, \\
K^{(2)} &= \sigma_x^{(2)} \otimes \sigma_z^{(1)} \otimes \sigma_z^{(3)}, \\
K^{(3)} &= \sigma_x^{(3)} \otimes \sigma_z^{(2)} \otimes I^{(1)} .
\end{aligned}
\end{equation*}

Let
\begin{equation*}
\begin{aligned}
|\phi\rangle_C =\,& \alpha |000\rangle + \beta |001\rangle + \gamma |010\rangle + \delta |011\rangle + \epsilon |100\rangle + \zeta |101\rangle + \eta |110\rangle + \theta |111\rangle .
\end{aligned}
\end{equation*}

Applying the stabilizer conditions $K^{(a)}|\phi\rangle_C = |\phi\rangle_C$ for $a=1,2,3$ leads to,
\begin{equation*}
\alpha = \beta = \gamma = \epsilon = -\delta = -\eta = -\zeta = -\theta .
\end{equation*}

Normalization yields $8|\alpha|^2 = 1$, hence $\alpha = \frac{1}{2\sqrt{2}}$. Therefore,
\begin{equation*}
\begin{aligned}
|\phi\rangle_C = \frac{1}{2\sqrt{2}} (&|000\rangle + |001\rangle + |010\rangle - |011\rangle + |100\rangle - |101\rangle - |110\rangle + |111\rangle).
\end{aligned}
\end{equation*}

This state admits the compact form,
\begin{equation}
|\phi\rangle_C = \frac{1}{\sqrt{2}} (|+0+\rangle + |-1-\rangle).
\end{equation}

\subsection*{Four Qubit Cluster State ($n=4$)}

For a four qubit linear cluster, the stabilizer operators are,
\begin{equation*}
\begin{aligned}
K^{(1)} &= \sigma_x^{(1)} \otimes \sigma_z^{(2)} \otimes I^{(3)} \otimes I^{(4)}, \\
K^{(2)} &= \sigma_x^{(2)} \otimes \sigma_z^{(1)} \otimes \sigma_z^{(3)} \otimes I^{(4)}, \\
K^{(3)} &= \sigma_x^{(3)} \otimes I^{(1)} \otimes \sigma_z^{(2)} \otimes \sigma_z^{(4)}, \\
K^{(4)} &= \sigma_x^{(4)} \otimes I^{(1)} \otimes I^{(2)} \otimes \sigma_z^{(3)} .
\end{aligned}
\end{equation*}

The resulting four-qubit cluster state can be written as
\begin{equation}
\begin{aligned}
|\phi\rangle_C = \frac{1}{2} (&|+\rangle_1 |0\rangle_2 |+\rangle_3 |0\rangle_4 
+ |+\rangle_1 |0\rangle_2 |-\rangle_3 |1\rangle_4 + |-\rangle_1 |1\rangle_2 |-\rangle_3 |0\rangle_4 + |-\rangle_1 |1\rangle_2 |+\rangle_3 |1\rangle_4 ).
\end{aligned}
\end{equation}

Equivalent representations follow from local basis changes, for example
\begin{equation}
\begin{aligned}
|\phi\rangle_C = \frac{1}{2} (&|0\rangle_1 |+\rangle_2 |0\rangle_3 |+\rangle_4
+ |0\rangle_1 |-\rangle_2 |1\rangle_3 |+\rangle_4 + |1\rangle_1 |-\rangle_2 |0\rangle_3 |+\rangle_4 + |1\rangle_1 |+\rangle_2 |1\rangle_3 |-\rangle_4 ).
\end{aligned}
\end{equation}

These explicit constructions illustrate how one dimensional cluster states arise naturally as stabilizer eigenstates and serve as the fundamental resource for measurement based quantum computation.

\subsection{Measurement Based Correction Patterns from Elementary Quantum Circuits}

It is essential to understand how the choice of measurement basis enables the implementation of arbitrary single qubit gates within MBQC from a circuit model perspective. Drawing an analogy with quantum teleportation, corrective operations are required to recover the desired output state depending on the measurement outcomes. Let $m \in \{0,1\}$ denote the measurement result of qubit~0. For $m=0$, the post measurement state is $|\psi\rangle = a|0\rangle + b|1\rangle$, whereas for $m=1$, the state becomes $|\psi\rangle = a|0\rangle - b|1\rangle$.

\begin{figure}[h!]
    \centering
    \includegraphics[width=0.9\textwidth,trim=20 20 20 20,clip]{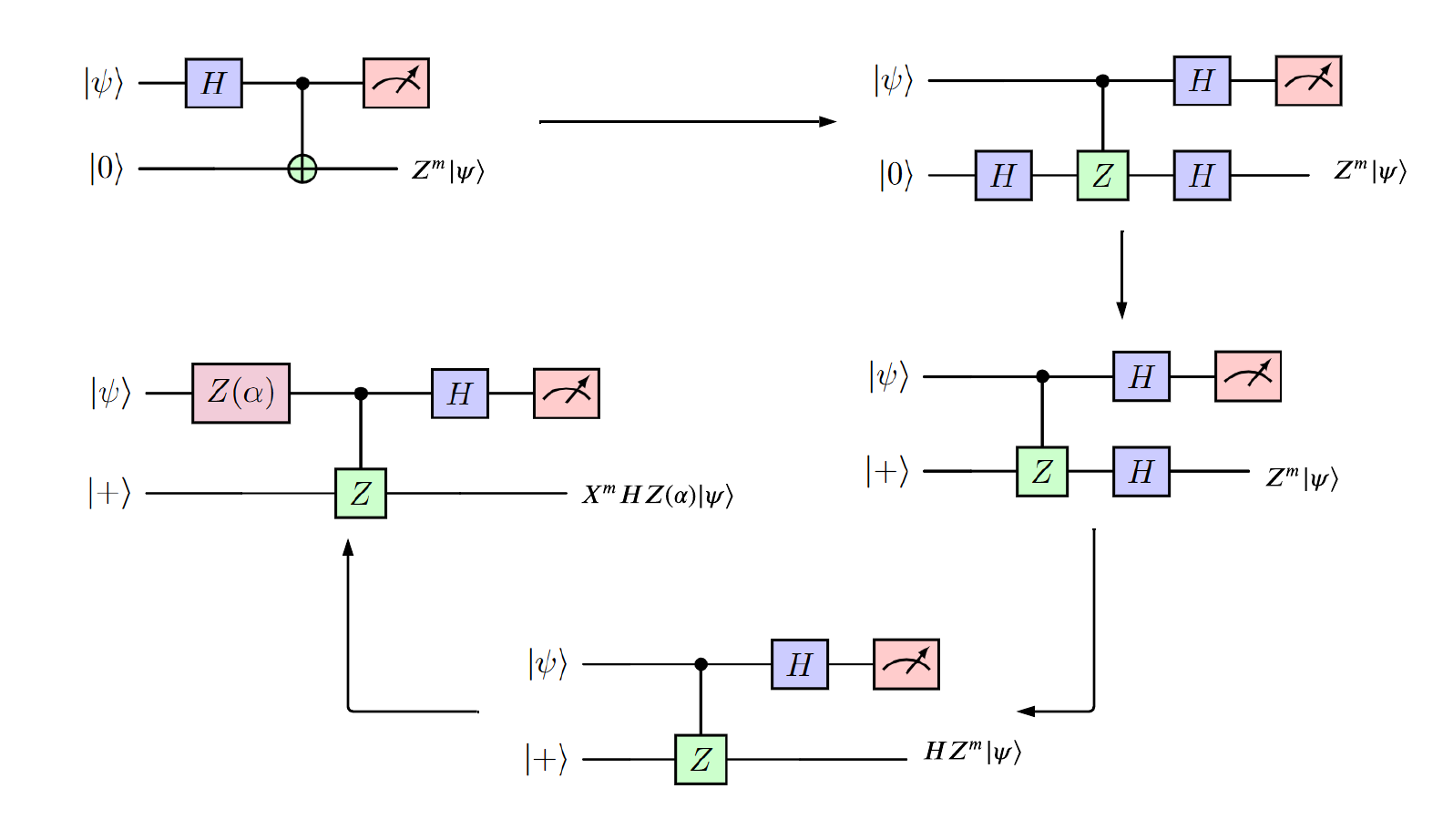}
    \caption{Derivation of measurement based correction patterns from elementary quantum circuits, illustrating how measurement outcomes in rotated bases lead to Pauli byproduct operators and adaptive corrections in MBQC.}
    \label{fig:MBQC_Circuit}
\end{figure}

\subsection{Measurement Pattern and Byproduct Correction}

In general, a graph (cluster) state associated with a graph $G = (V,E)$ is defined as,
\begin{equation}
\ket{G} = \prod_{(i,j)\in E} CZ_{i,j} \bigotimes_{i\in V} \ket{+}_i ,
\end{equation}
where $CZ_{i,j}$ denotes the controlled-$Z$ operation acting on vertices $i$ and $j$, and,
\begin{equation*}
\ket{+} = \frac{1}{\sqrt{2}}(\ket{0}+\ket{1}).
\end{equation*}

The tensor product $\bigotimes_{i\in V} \ket{+}_i$ explicitly denotes,
\begin{equation*}
\ket{+}_{i_1} \otimes \ket{+}_{i_2} \otimes \ket{+}_{i_3} \otimes \cdots .
\end{equation*}

\subsection*{Single Qubit State Transfer via a Two Qubit Cluster}

Consider the simplest nontrivial cluster consisting of two qubits labeled $0$ and $1$.  
Let an arbitrary input state $\ket{\psi_{\mathrm{in}}}$ be encoded on qubit $0$, while qubit $1$ is prepared in $\ket{+}$.  
The joint state after entanglement is,
\begin{equation*}
\ket{g'} = CZ_{0,1} \left( \ket{\psi_{\mathrm{in}}}_0 \otimes \ket{+}_1 \right).
\end{equation*}

Measurement of qubit $0$ is performed in the $X$ basis, with projectors,
\begin{equation*}
P_{X,s} =
\begin{cases}
\ket{+}\!\bra{+}, & s=0, \\
\ket{-}\!\bra{-}, & s=1,
\end{cases}
\end{equation*}
where $\ket{-} = (\ket{0}-\ket{1})/\sqrt{2}$.

Conditioned on the measurement outcome $s_0 \in \{0,1\}$, the post-measurement state on qubit $1$ becomes,
\begin{equation*}
\ket{g_{s_0}} = \bra{\pm}_{0} \, CZ_{0,1} \left( \ket{\psi_{\mathrm{in}}}_0 \otimes \ket{+}_1 \right),
\end{equation*}
which evaluates to,
\begin{equation*}
\ket{g_{s_0}} =
\begin{cases}
H \ket{\psi_{\mathrm{in}}}_1, & s_0 = 0, \\
X H \ket{\psi_{\mathrm{in}}}_1, & s_0 = 1.
\end{cases}
\end{equation*}

Thus, the input state is transferred to qubit $1$ up to a Pauli byproduct operator,
\begin{equation}
\ket{\psi_{\mathrm{in}}}_1 = X^{s_0} \, \bra{\pm}_0 \, CZ_{0,1} \left( \ket{\psi_{\mathrm{in}}}_0 \otimes \ket{+}_1 \right).
\end{equation}

\subsection*{Single Qubit Rotation $R_Z(\theta)$ via Measurement}

A single qubit $Z$-rotation can be implemented by modifying the measurement basis.  
Applying a local rotation $R_Z(\theta)$ before measurement yields
\begin{equation}
R_Z(\theta)\ket{\psi_{\mathrm{in}}}_1
= X^{s_0} \bra{\pm_{-\theta}}_0 \, CZ_{0,1}
\left( \ket{\psi_{\mathrm{in}}}_0 \otimes \ket{+}_1 \right).
\end{equation}

Extending this to a linear three qubit cluster enables the realization of arbitrary single-qubit unitaries.  
Specifically,
\begin{equation}
\begin{aligned}
R_Z(\theta)\ket{\psi_{\mathrm{in}}}_2
= Z_2^{s_0} X_2^{s_1} \,
\bra{\pm}_1 \bra{\pm_{-\theta}}_0 \,
CZ_{1,2} CZ_{0,1}
\ket{+}_2 \ket{+}_1 \ket{\psi_{\mathrm{in}}}_0 ,
\end{aligned}
\end{equation}
where the Pauli byproduct operators depend explicitly on the classical measurement outcomes
$s_0$ and $s_1$.

\subsection*{Two-Qubit Gate: Measurement-Based CNOT}

Measurement-based quantum computation also allows the deterministic implementation of entangling gates.  
For a suitable four-qubit cluster configuration, a CNOT gate between logical qubits $0$ (control) and $3$ (target) is obtained as
\begin{equation}
\begin{aligned}
\mathrm{CNOT}_{0,3}\ket{\psi_{\mathrm{in}}}_{03}
= X_3^{s_2} Z_3^{s_1} Z_0^{s_1} \,
\bra{\pm}_2 \bra{\pm}_1 \,
CZ_{0,2} CZ_{2,3} CZ_{1,2} \,
\ket{+}_3 \ket{+}_2 \ket{\psi_{\mathrm{in}}}_{01}.
\end{aligned}
\end{equation}

Here, all non-determinism introduced by measurements is absorbed into known Pauli byproduct operators, which can be corrected either actively or via classical feedforward.

\subsection{MBQC Circuit for Different Gate like functionality}

\begin{figure}[t]
    \centering

    \begin{subfigure}[t]{0.48\textwidth}
        \centering
        \includegraphics[width=\linewidth]{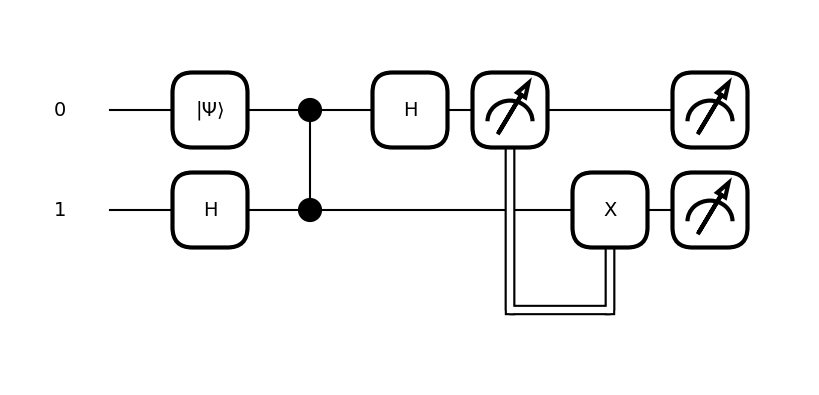}
        \caption{}
        \label{fig:mbqc_h}
    \end{subfigure}
    \hfill
    \begin{subfigure}[t]{0.48\textwidth}
        \centering
        \includegraphics[width=\linewidth]{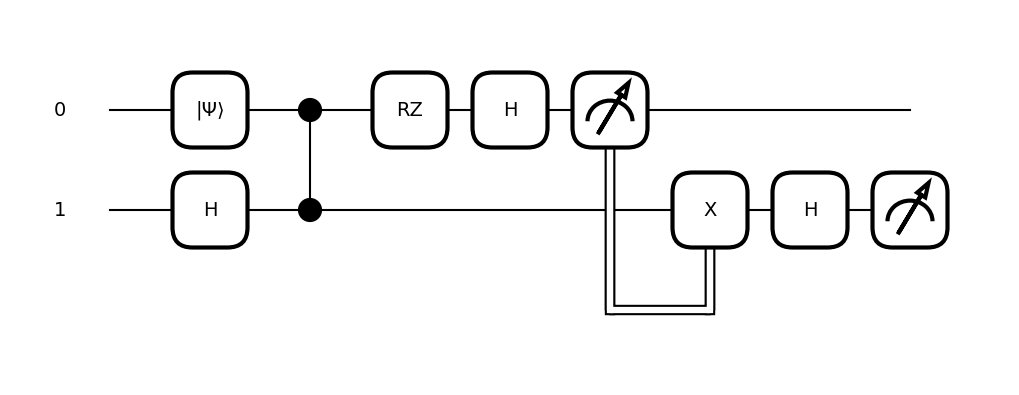}
        \caption{}
        \label{fig:mbqc_two_qubit_rot}
    \end{subfigure}

    \vspace{1em}

    \begin{subfigure}[t]{0.48\textwidth}
        \centering
        \includegraphics[width=\linewidth]{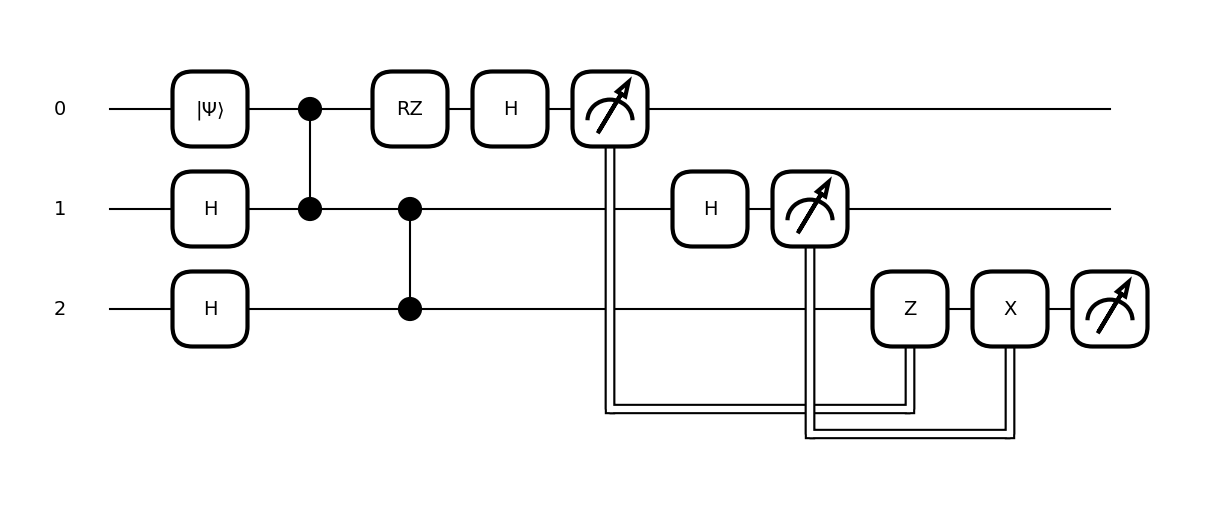}
        \caption{}
        \label{fig:mbqc_rz}
    \end{subfigure}
    \hfill
    \begin{subfigure}[t]{0.48\textwidth}
        \centering
        \includegraphics[width=\linewidth]{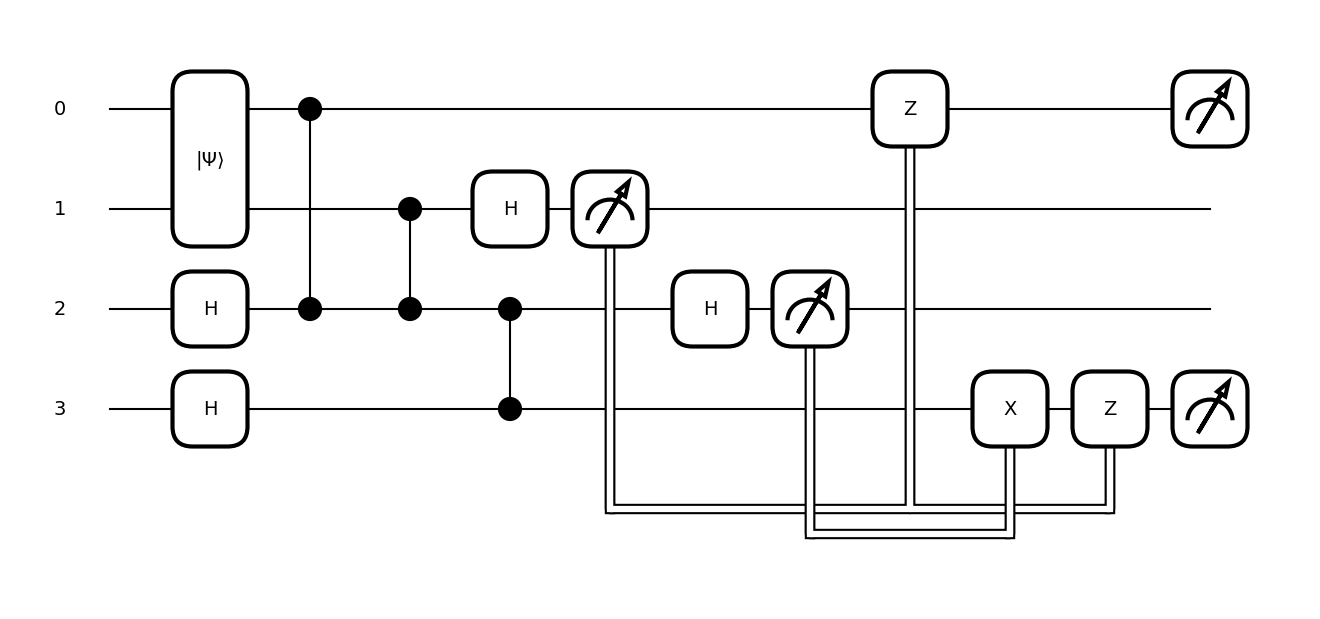}
        \caption{}
        \label{fig:mbqc_cnot}
    \end{subfigure}

    \caption{Measurement-based quantum computation (MBQC) gate constructions: 
    (a) Hadamard gate implemented using a two-qubit linear cluster; 
    (b) single-qubit state transfer and rotation via a two-qubit cluster; 
    (c) measurement-based realization of the $R_Z(\theta)$ gate using a three-qubit cluster; 
    (d) CNOT-equivalent entangling gate implemented with a multi-qubit cluster. 
    Logical operations are achieved through single-qubit measurements and classical feedforward, with deterministic behavior up to Pauli byproduct operators.}
    \label{fig:mbqc_gates}
\end{figure}

\begin{figure}[h!]
    \centering
    \begin{subfigure}{0.32\textwidth}
        \centering
        \includegraphics[width=\linewidth]{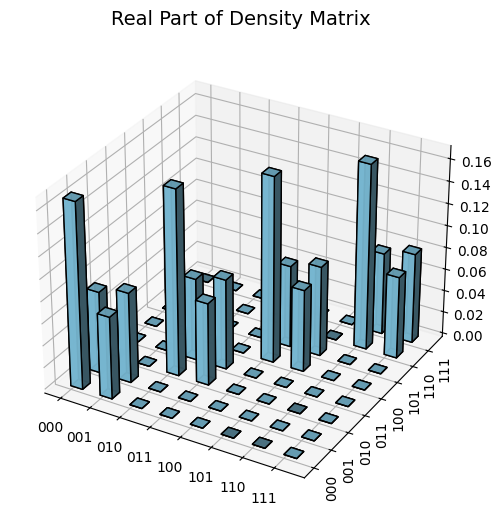}
        \label{fig:S_gate}
    \end{subfigure}
    \hfill
    \begin{subfigure}{0.32\textwidth}
        \centering
        \includegraphics[width=\linewidth]{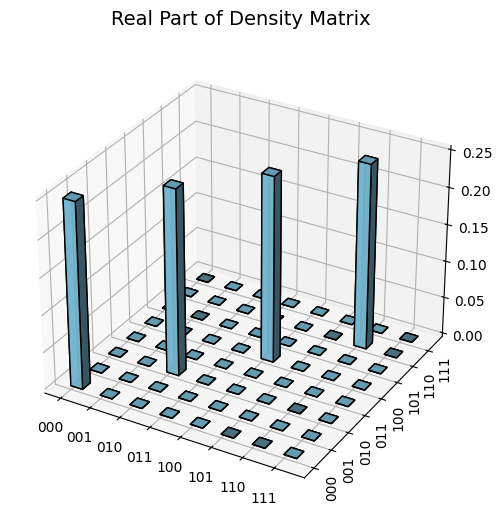}
        \label{fig:T_gate}
    \end{subfigure}
    \hfill
    \begin{subfigure}{0.32\textwidth}
        \centering
        \includegraphics[width=\linewidth]{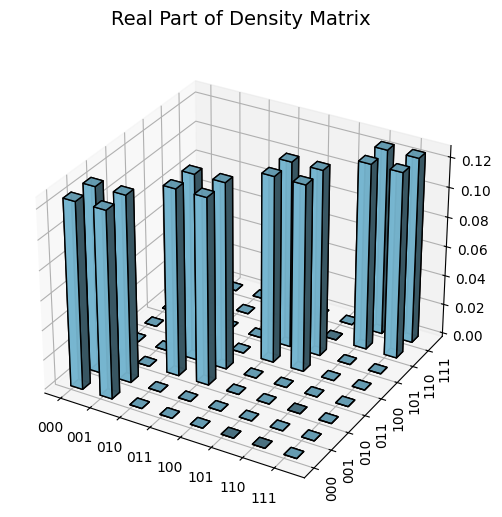}
        \label{fig:H_gate}
    \end{subfigure}

    \caption{Density matrices for three‑qubit cluster MBQC circuits:  
    (a) $S$ gate;  
    (b) $T$ gate;  
    (c) $H$ gate.}
    \label{fig:S_T_H_gates}
\end{figure}

Measurement‑based quantum computation (MBQC) circuits can reproduce the functionality of standard quantum gates by exploiting entangled cluster states and adaptive measurements. Different gate operations—such as single‑qubit rotations or controlled‑Z interactions are realized through specific measurement bases, with classical correction patterns applied depending on the measurement outcomes. This framework highlights how MBQC provides gate‑like behavior without direct unitary evolution, instead relying on entanglement and measurement to drive computation. MBQC circuits are shown in Fig \ref{fig:mbqc_gates}. We can also choose different rotation to get the gate like functionality from two and three qubit cluster states shown in Table \ref{tab:gate_rotation}. Also to check the validity we can check density matrix of the MBQC circuit which is shown in Fig \ref{fig:S_T_H_gates}.

\begin{table}[h!]
\centering
\caption{Measurement-based quantum computation (MBQC) implementation of single-qubit $Z$-rotations. Each gate is realized by choosing an appropriate rotation angle and measurement basis.}
\vspace{0.3cm} 
\begin{tabular}{c c c}
\toprule
Gate & Rotation Angle & Measurement Basis \\
\midrule
$H$ & $0$ & $M(\phi)$ in $XY$-plane\\
$S$ & $\pi/2$ & $M(\phi)$ in $XY$-plane \\
$T$ & $\pi/4$ & $M(\phi)$ in $XY$-plane \\
$Z$ & $\pi$   & Pauli-$Z$ basis \\
\bottomrule
\end{tabular}
\label{tab:gate_rotation}
\end{table}

\subsection{Mapping Ideal Oracle to MBQC Flow}


\end{document}